\documentclass[12pt]{article}
\usepackage{epsfig,cite}
\usepackage {amsmath}
\usepackage{amssymb}
\include{epsf}
\newcount\timecount
\newcount\hours \newcount\minutes  \newcount\temp \newcount\pmhours
\hours = \time
\divide\hours by 60
\temp = \hours
\multiply\temp by 60
\minutes = \time
\advance\minutes by -\temp
\def\hour{\the\hours}
\def\minute{\ifnum\minutes<10 0\the\minutes
            \else\the\minutes\fi}
\def\clock{
\ifnum\hours=0 12:\minute\ AM
\else\ifnum\hours<12 \hour:\minute\ AM
      \else\ifnum\hours=12 12:\minute\ PM
            \else\ifnum\hours>12
                 \pmhours=\hours
                 \advance\pmhours by -12
                 \the\pmhours:\minute\ PM
                 \fi
            \fi
      \fi
\fi
}

\def\monthname{\relax\ifcase\month 0/\or January\or February\or
   March\or April\or May\or June\or July\or August\or September\or
   October\or November\or December\else\number\month/\fi}

\def\bold#1{\setbox0=\hbox{$#1$}%
     \kern-.025em\copy0\kern-\wd0
     \kern.05em\copy0\kern-\wd0
     \kern-.025em\raise.0433em\box0 }

\def\beq{\begin{equation}}
\def\eeq{\end{equation}}

\def\ga{\mathrel{\raise.3ex\hbox{$>$\kern-.75em\lower1ex\hbox{$\sim$}}}}
\def\la{\mathrel{\raise.3ex\hbox{$<$\kern-.75em\lower1ex\hbox{$\sim$}}}}
\def\gev{{\rm \, Ge\kern-0.125em V}}
\def\tev{{\rm \, Te\kern-0.125em V}}
\def\gyr{{\rm \, G\kern-0.125em yr}}

\def\gappeq{\mathrel{\rlap {\raise.5ex\hbox{$>$}}
{\lower.5ex\hbox{$\sim$}}}}
\def\lappeq{\mathrel{\rlap{\raise.5ex\hbox{$<$}}
{\lower.5ex\hbox{$\sim$}}}}
\def\Toprel#1\over#2{\mathrel{\mathop{#2}\limits^{#1}}}

\def\m12{m_{1\!/2}}

\def\PL{{Phys.~Lett.} }

\def\bea{\begin{eqnarray}}
\def\eea{\end{eqnarray}}

\begin{document}
\begin{titlepage}
\pagestyle{empty}
\baselineskip=21pt
\rightline{CERN-PH-TH/2005-173, UMN--TH--2416/05, FTPI--MINN--05/43, ACT-09-05, MIFP-05-23}
\vskip 0.1in
\begin{center}
{\large{\bf On the Higgs Mass in the CMSSM}}
\end{center}
\begin{center}
\vskip 0.1in
{\bf John~Ellis}$^1$, 
{\bf Dimitri~Nanopoulos}$^2$, 
{\bf Keith~A.~Olive}$^3$
and {\bf Yudi~Santoso}$^4$
\vskip 0.1in
{\small {\it
$^1${TH Division, PH Department, CERN, Geneva, Switzerland}\\
$^2${George P. and Cynthia W. Mitchell Institute for Fundamental
Physics, \\ Texas A\&M
University, College Station, TX 77843, USA; \\
Astroparticle Physics Group, Houston
Advanced Research Center (HARC),
Mitchell Campus,
Woodlands, TX~77381, USA; \\
Academy of Athens,
Division of Natural Sciences, 28~Panepistimiou Avenue, Athens 10679,
Greece}\\
$^3${William I. Fine Theoretical Physics Institute, \\
University of Minnesota, Minneapolis, MN 55455, USA}\\
$^4${Department of Physics and Astronomy, University of Victoria, \\
Victoria, BC, V8P 1A1, Canada; \\
Perimeter Institute of Theoretical Physics, Waterloo, ON, N2J 2W9, Canada}\\
}}

\vskip 0.2in
{\bf Abstract}
\end{center}
\baselineskip=18pt \noindent

We estimate the mass of the lightest neutral Higgs boson $h$ in the
minimal supersymmetric extension of the Standard Model with universal soft
supersymmetry-breaking masses (CMSSM), subject to the available
accelerator and astrophysical constraints. For $m_t = 174.3$~GeV, we find
that $114~{\rm GeV} < m_h < 127$~GeV and a peak in the $\tan \beta$
distribution $\simeq 55$. We observe two distinct peaks in the
distribution of $m_h$ values, corresponding to two different regions of
the CMSSM parameter space. Values of $m_h < 119$~GeV correspond to small
values of the gaugino mass $m_{1/2}$ and the soft trilinear
supersymmetry-breaking parameter $A_0$, lying along coannihilation strips,
and most of the allowed parameter sets are consistent with a
supersymmetric interpretation of the possibly discrepancy in $g_\mu - 2$.
On the other hand, values of $m_h > 119$~GeV may correspond to much larger
values of $m_{1/2}$ and $A_0$, lying in rapid-annihilation funnels. The
favoured ranges of $m_h$ vary with $m_t$, the two peaks being more clearly
separated for $m_t = 178$~GeV and merging for $m_t = 172.7$~GeV. If the
$g_\mu - 2$ constraint is imposed, the mode of the $m_h$ distribution is
quite stable, being $\sim 117$~GeV for all the studied values of $m_t$.

\vfill
\leftline{CERN-PH-TH/2005-173}
\leftline{September 2005}
\end{titlepage}
\baselineskip=18pt

\section{Introduction}

One of the characteristic predictions of the minimal supersymmetric  extension
of the Standard Model (MSSM) is the mass of the lightest neutral  Higgs boson
$h$, which is expected to be $m_h \la 150$~GeV~\cite{mssmhiggs}. This is very consistent
with the range $m_h \la 200$~GeV that is favoured by global  analyses of the
available precision electroweak data~\cite{lepewwg}. Various studies have shown
that the lightest neutral MSSM Higgs boson is very likely to be  discovered at
the LHC, and possibly at the Fermilab Tevatron 
collider. It is therefore interesting to attempt to refine the MSSM 
prediction for $m_h$, and to consider what one would learn from a  measurement
of the $h$ mass \cite{egno}.

We study these questions within the constrained version of the MSSM
(CMSSM), in which the soft supersymmetry-breaking scalar masses $m_0$ and
gaugino masses $m_{1/2}$ are each assumed to be universal at some GUT
input scale, as are the trilinear soft supersymmetry-breaking parameters
$A_0$. We impose on the CMSSM the available phenomenological constraints
from accelerator experiments, astrophysics and cosmology \cite{EHNOS,cmssmwmap}, treating the
supersymmetric interpretation of the anomalous magnetic moment of the
muon, $g_\mu - 2$, as an optional constraint, and interpreting the WMAP
range for the cold dark matter density~\cite{WMAP} as an upper bound: 
$\Omega_\chi h^2 < 0.129$.

\begin{figure}
\vskip 0.5in
\vspace*{-0.75in}
\begin{minipage}{8in}
\epsfig{file=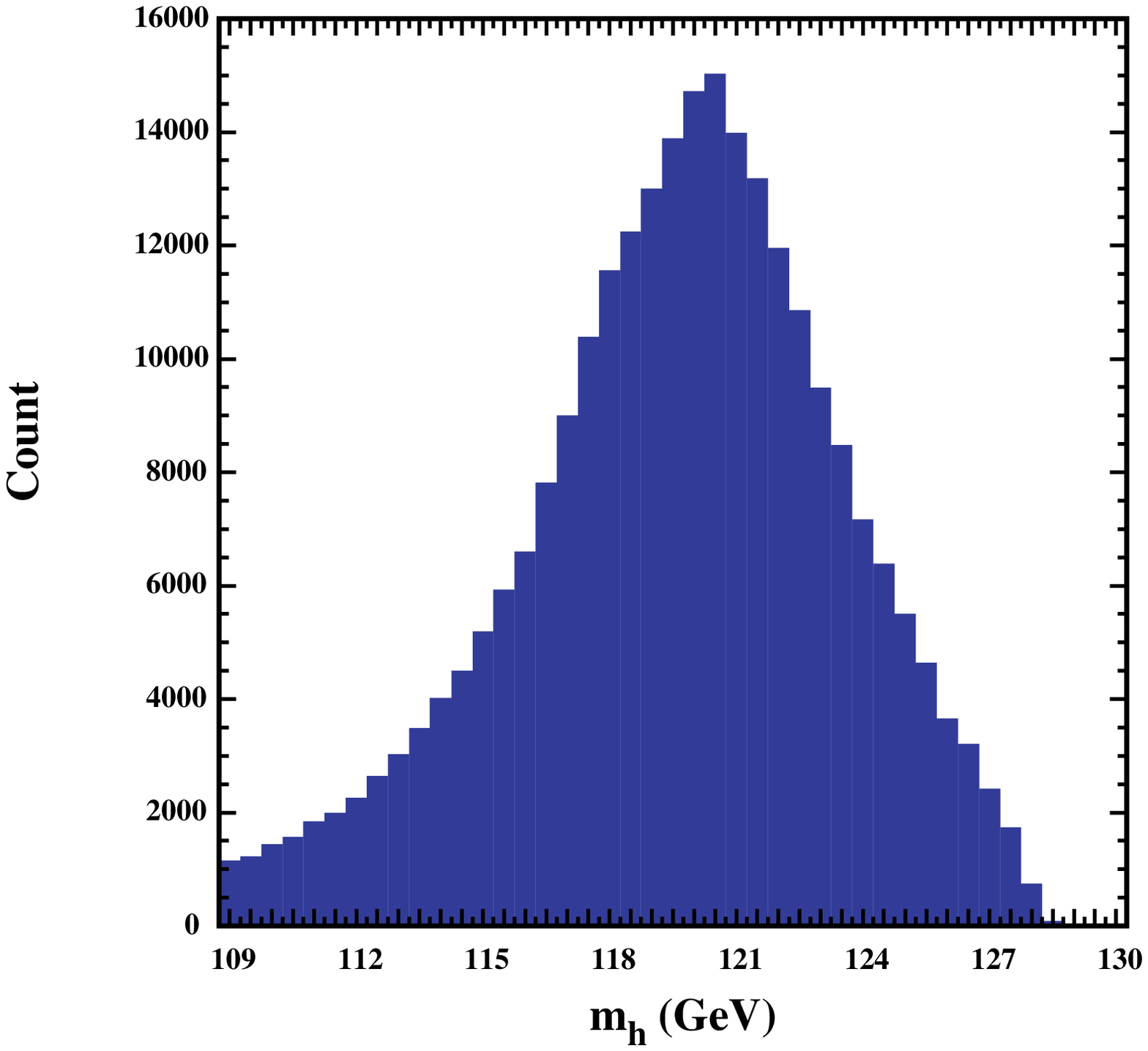,height=3.1in}
\epsfig{file=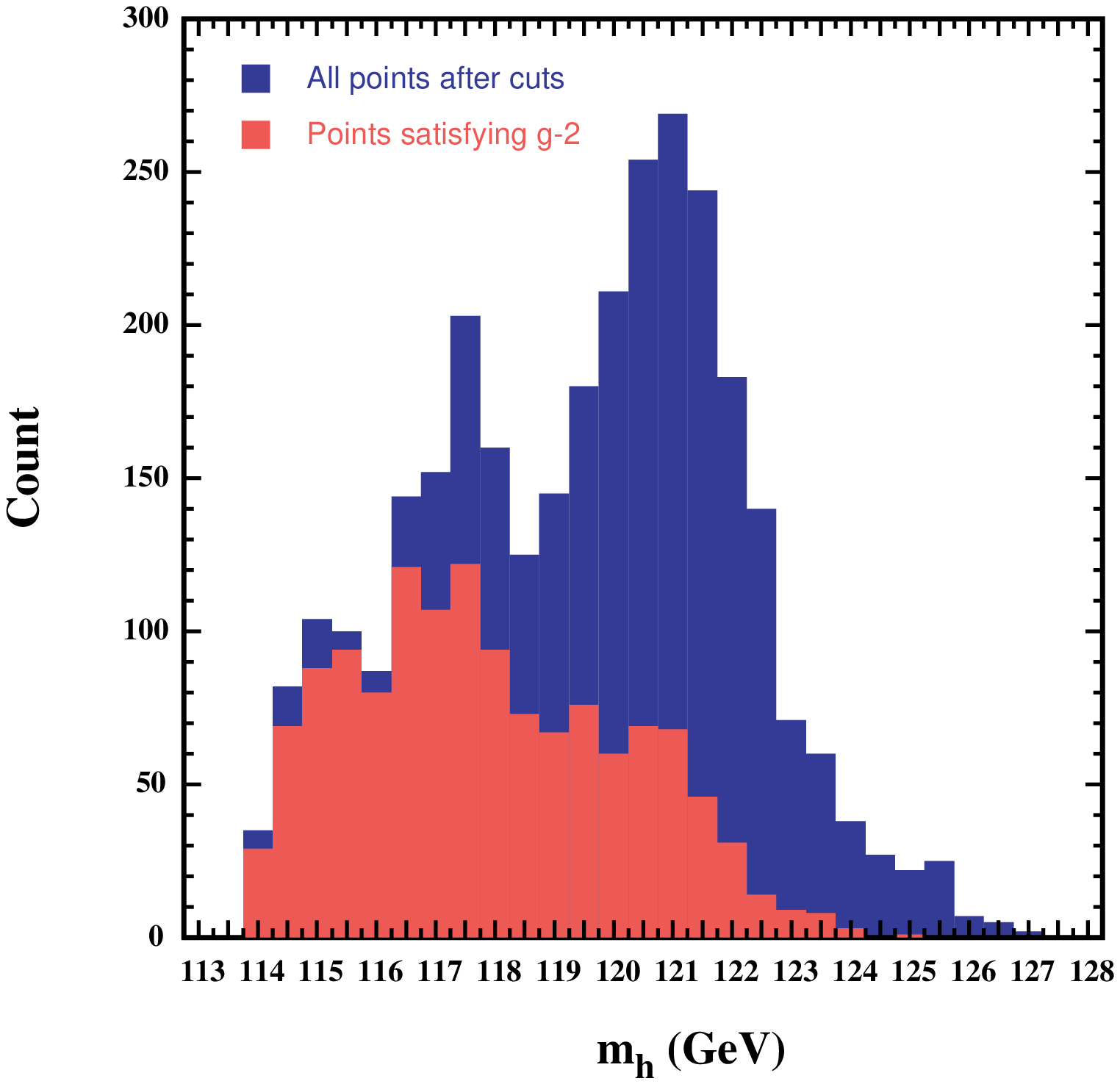,height=3.1in}
\hfill
\end{minipage}
\caption{
{\it
The distribution of the mass of the lightest neutral Higgs 
boson, $m_h$, in the CMSSM (a) before applying the accelerator cuts and 
the WMAP relic density constraint, and (b) after applying these 
constraints.
In the latter case, the red (light) histogram shows the points favoured by 
the optional $g_\mu - 2$ constraint. We assume $m_t = 174.3$~GeV in both 
panels.
}} 
\label{fig:mh}
\end{figure}

We base our study on a statistical sampling of the CMSSM parameter space
that is uniform in the $(m_{1/2}, m_0)$ plane for 100~GeV $< m_{1/2} <
2$~TeV, $m_0 < 2$~TeV, $|A_0 / m_{1/2}| < 3$, $2 < \tan \beta < 58$ and
$\mu > 0$, assuming initially that $m_t = 174.3$~GeV~\cite{mtop174} and
discussing later
other possible values of $m_t$.  We began with a random sample of over
320,000 CMSSM points: requiring the lightest supersymmetric particle (LSP)
to be a neutralino brought the number down to somewhat over 260,000. As
seen in Fig.~\ref{fig:mh}(a), before imposing the various phenomenological
constraints we find that $m_h$~\footnote{We use Fortran code {\tt
FeynHiggs}~\cite{feynhiggs} to calculate $m_h$.} 
is distributed between very low values $ <
110$~GeV and an upper limit $\sim 128$~GeV, with a single peak at $\sim
120$~GeV.  The drop-off in the count at low $m_h$ is mainly related to our
choice of a uniform measure in the CMSSM input parameters: because of the
logarithmic dependence of $m_h$ on $m_{1/2}$ and $m_0$, low values of
$m_h$ only occur at low values of $m_{1/2}, m_0$ and $\tan \beta$. (We
recall that $m_h$ evolves quickly as $m_{1/2}$ is increased at low
$m_{1/2}$ and more slowly at large $m_{1/2}$.)  The fall-off at large
$m_h$ is largely due to our choice of 2~TeV as the upper limit on
$m_{1/2}$. Extending this upper limit would slowly push the peak in
Fig.~\ref{fig:mh}(a) to the right, and the count at the peak would grow
rapidly. Once again, because of the logarithmic dependence of $m_h$ on
$m_{1/2}$, even a modest change in the position of the peak would require
increasing the upper limit on $m_{1/2}$ substantially.

We next apply a series of constraints, including the LEP lower limit on 
the
chargino mass of 104~GeV, $b \to s \gamma$~\cite{bsg} and the limit on
$\Omega_\chi h^2$ that was discussed above, as well as the the direct
experimental limit on $m_h$ of $\sim 114$~GeV~\cite{mh}~\footnote{We recall
that, although this lower limit may be relaxed in some variants of the MSSM, its
value does not change for the CMSSM studied here~\cite{nuhm1}.}.  The most
severe cut on the sample, by far, is that due to the relic density, which for
most points exceeds the WMAP upper limit. When all cuts are applied our sample
is reduced to 3075 points, which are plotted in Fig.~\ref{fig:mh}(b).  
Most of
the range in $m_h$ is still available after imposing the various
phenomenological constraints, as seen in Fig.~\ref{fig:mh}(b).  However, we see
that the distribution of $m_h$ within this range exhibits significant
structures, with peaks at $m_h \simeq 121$ and $117$~GeV, and a dip at
$m_h \simeq 119$~GeV.

In the rest of this paper, we explain the origins of these features,
describe the domains of the CMSSM parameter space that populate these
peaks in the $m_h$ distribution, and discuss the effects of imposing the
optional $g_\mu - 2$ constraint~\cite{g-2} and varying $m_t$. The peaked
structures
in $\tan \beta$ and $m_h$ reflect different processes that might reduce
the density of supersymmetric relics $\chi$ into the range allowed by WMAP
and other observations~\footnote{Note that although
we consider $A_0 \ne 0$, the the stop-coannihilation 
region \cite{stopco} is beyond the range we scan.}: either coannihilations with sleptons, most
importantly the lighter stau: ${\tilde \tau}_1$ \cite{efo}, or rapid 
annihilations:  $\chi \chi$ via the
heavier neutral Higgs bosons $A,H$ \cite{funnel}, or (exceptionally) rapid
annihilations: $\chi \chi$ via the lightest neutral Higgs bosons 
$h$~\cite{Drees}.

The structures in the $m_h$ distribution imply that, once $m_t$ is better
known from Tevatron and/or LHC measurements and assuming the CMSSM
framework~\footnote{We also assume that theoretical errors in the CMSSM
calculation of $m_h$ can be reduced along with the experimental error.}, a
measurement of $m_h$ at the LHC or Tevatron might enable one to estimate
ranges for the values of $m_{1/2}, A_0$ and $\tan \beta$, {\it even if
sparticles themselves are not yet discovered}. If sparticles {\it are}
discovered, confronting their masses with the ranges inferred from $m_h$
will be a crucial test of the CMSSM framework.

\section{Effects of Phenomenological Constraints on the CMSSM Parameter 
Space}

\begin{figure}
\vskip 0.5in
\vspace*{-0.75in}
\begin{minipage}{8in}
\epsfig{file=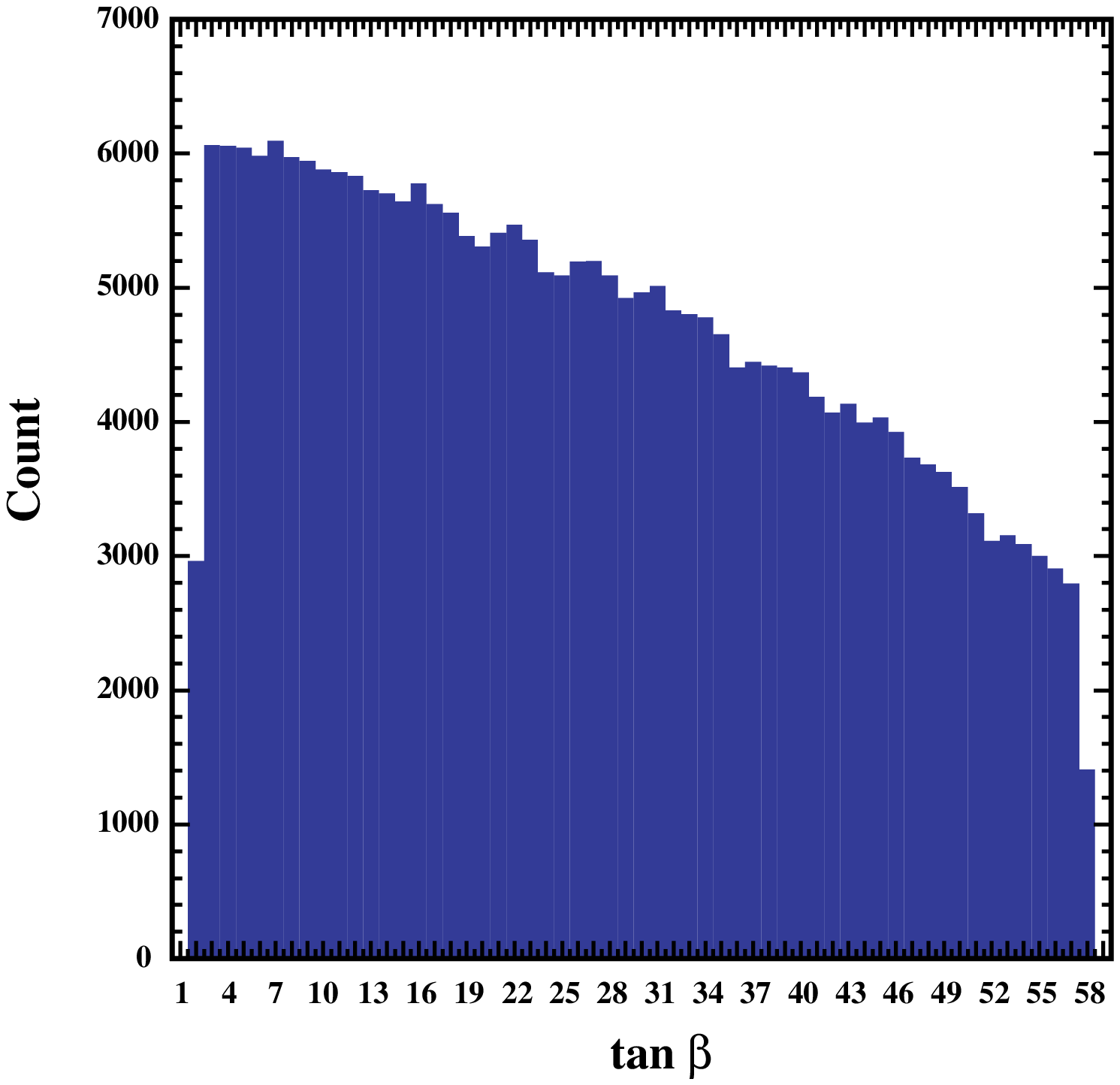,height=3.1in}
\epsfig{file=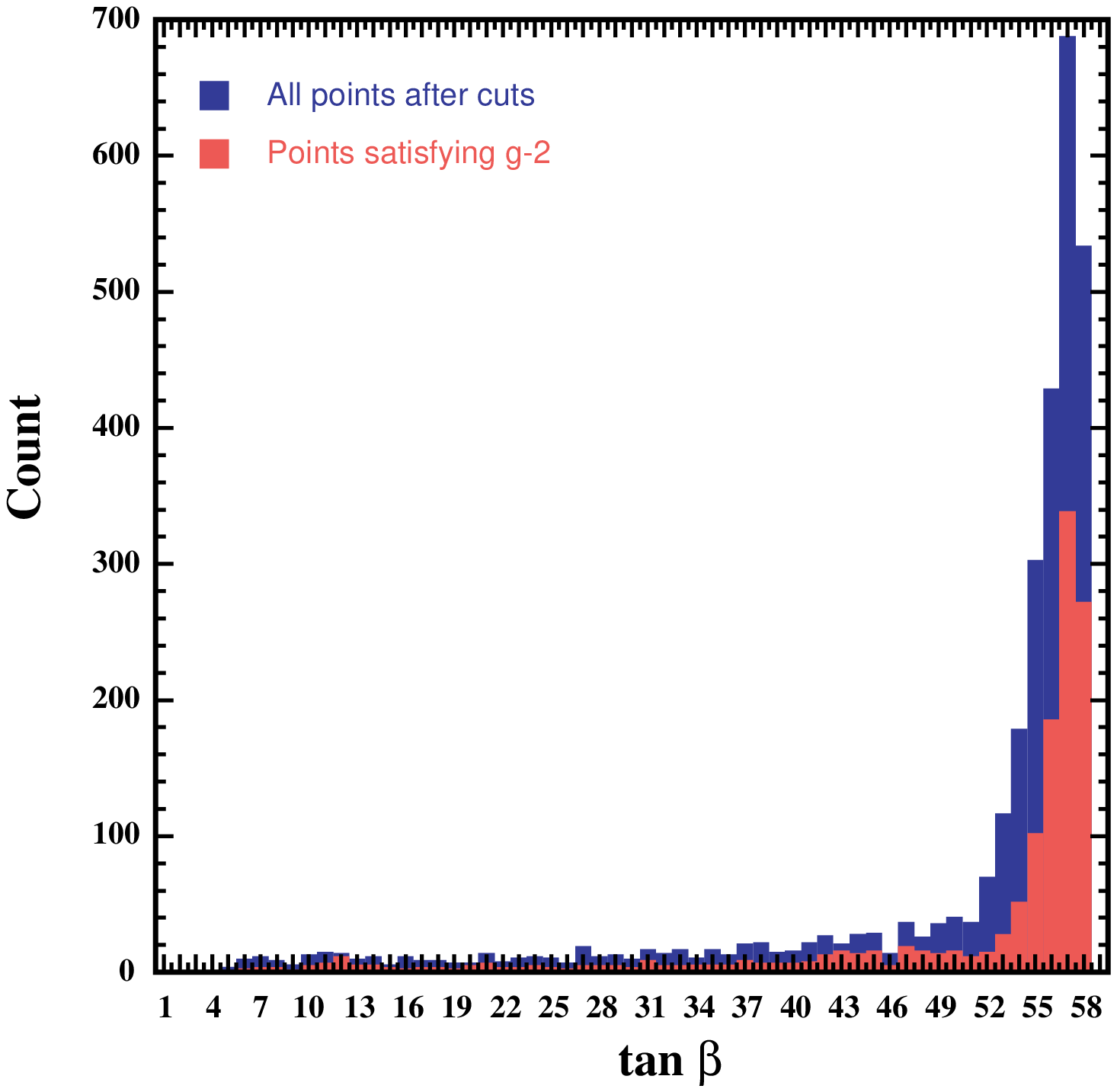,height=3.1in}
\hfill
\end{minipage}
\caption{
{\it
(a) The distribution in $\tan \beta$ of the sample points shown in 
Fig.~\protect\ref{fig:mh}(a), before the accelerator and 
WMAP constraints are applied, and (b) their distribution after applying 
these phenomenological constraints. In the latter, we see that the 
distribution of $g_\mu - 2$-friendly points (coloured red/grey) is similar 
to 
that of the total sample.
}} 
\label{fig:tanbeta}
\end{figure}

As already mentioned, we have sampled uniformly the $(m_{1/2}, m_0)$ plane
for $m_{1/2}, m_0 < 2$~TeV, $|A_0 / m_{1/2}| < 3$, $2 < \tan \beta < 58$
and $\mu > 0$~\footnote{This sign of $\mu$ is suggested by even a loose
interpretation of $g_\mu - 2$.}, assuming $m_t = 174.3$~GeV as our
default~\footnote{We discuss later the effect of varying $m_t$ in our
analysis.}. As $\tan \beta$ is increased, there is an increasing fraction
of sample points that do not yield consistent electroweak vacua.
Nevertheless, the consistent solutions are distributed quite smoothly in
$\tan \beta$ before applying the accelerator and cosmological cuts, as
seen in panel (a) of Fig.~\ref{fig:tanbeta}. However, after applying the
cuts, the distribution in $\tan \beta$ is far from uniform, as seen in
panel (b) of Fig.~\ref{fig:tanbeta}. The distribution of allowed models is
sharply peaked towards large $\tan \beta$, with a relatively small tail
surviving below $\tan \beta \sim 20$. This observation holds for both
the general sample and the $(g_\mu - 2)$-friendly subsample~\footnote{We assume
here $g_\mu-2$ range from 6.8 to $43.6 \times 10^{-10}$~\cite{g-2}.}, 
shown as the light (red) shaded histogram in panel (b).

The preference for large $\tan \beta$ is an understandable consequence of
the interplay of the various accelerator and cosmological constraints. For
example, the cosmological upper limit on the supersymmetric relic density
in the coannihilation region imposes an upper limit on $m_{1/2}$ that is
significantly relaxed at large $\tan \beta$, in particular by rapid $\chi
\chi \to A, H$ annihilation.  Moreover, the funnels due to the
rapid-annihilation processes $\chi \chi \to H, A$ are broader than the
coannihilation strips that define the acceptable cosmological regions at
lower $\tan \beta$. We also note that the range of small values of
$m_{1/2}$ that is excluded by the experimental lower limit on $m_h$
diminishes as $\tan \beta$ increases, and recall that the predominance of
high $\tan \beta$ in satisfying constraints was clearly seen in a
likelihood analysis~\cite{eossl} when comparing
regions of high likelihood for $\tan \beta = 10$ and 50.

\begin{figure}
\vskip 0.5in  
\vspace*{-0.75in}
\begin{minipage}{8in}
\epsfig{file=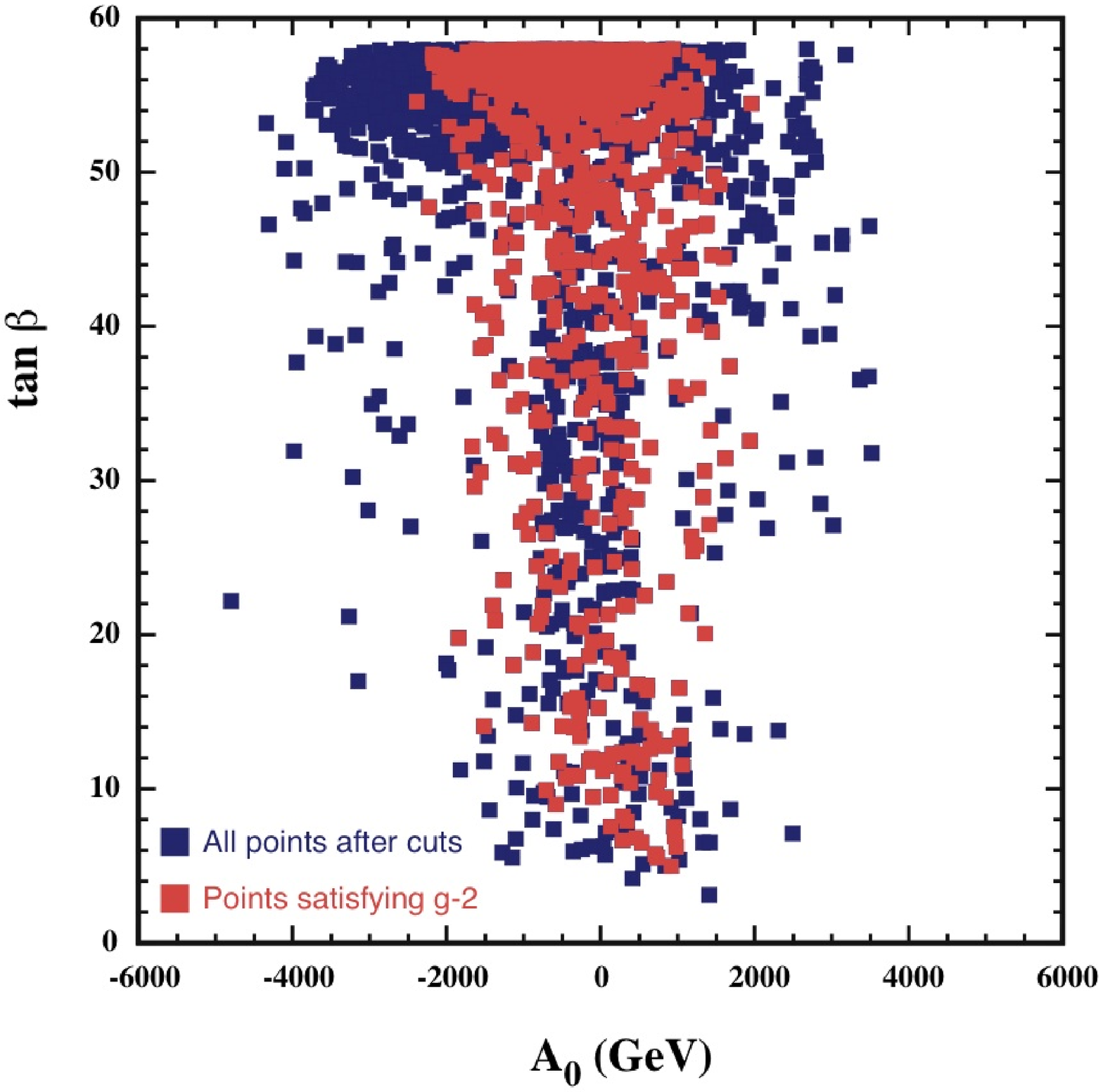,height=3.1in}
\epsfig{file=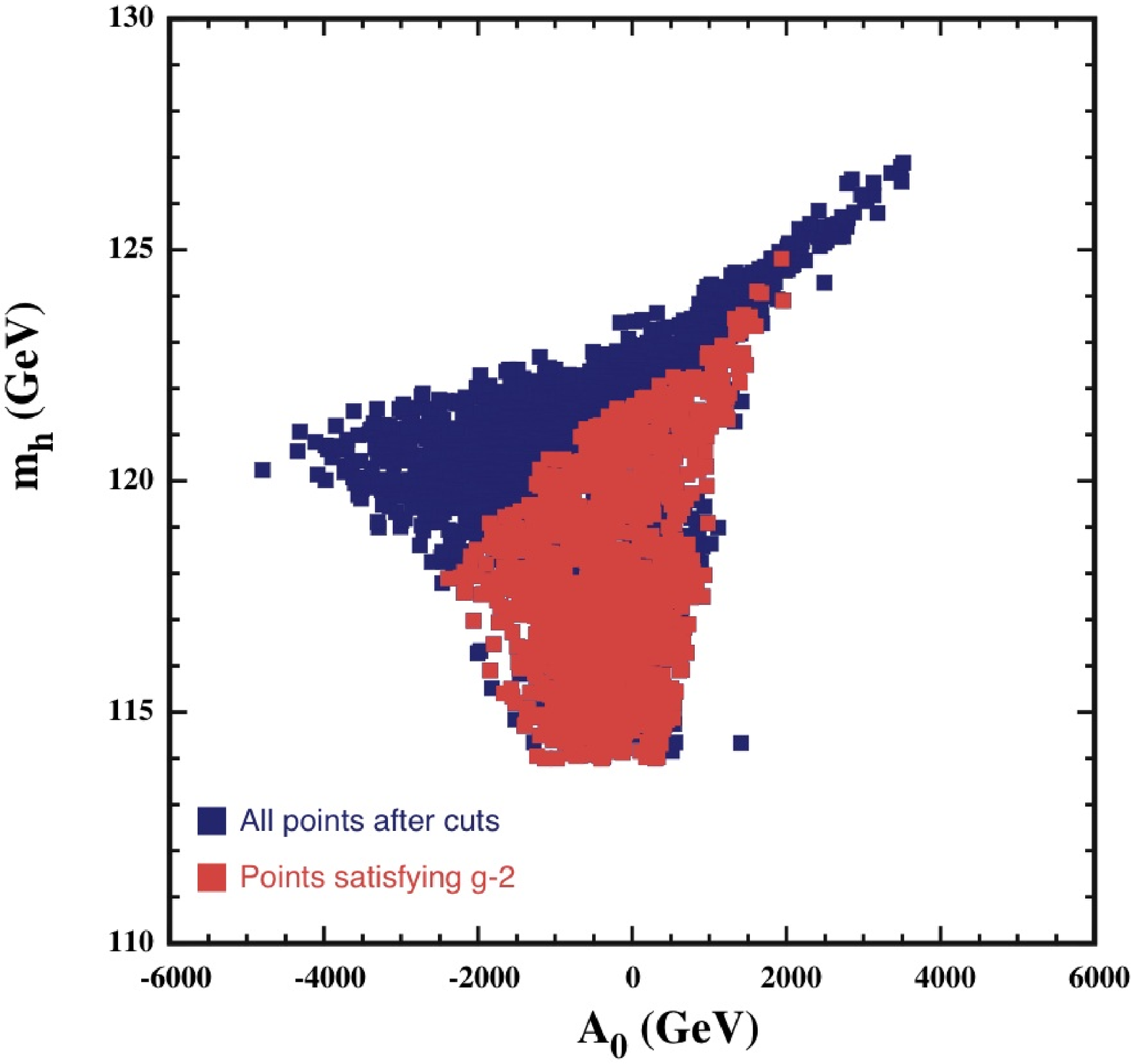,height=3.1in}
\hfill
\end{minipage}
\caption{
{\it
(a) The distribution in the $(A_0, \tan \beta)$ plane of the sample points 
shown in Fig.~\protect\ref{fig:mh}(b) that survive the accelerator and
WMAP constraints, and (b) a scatter plot of these points in the 
$(A_0, m_h)$ plane. The $(g_\mu - 2)$-friendly points are coloured red 
(grey).
}}
\label{fig:gen}
\end{figure}

Panel (a) of Fig.~\ref{fig:gen} displays the allowed points in the $(A_0,
\tan \beta)$ plane. We see that they gather into three clusters: one
centered around $A_0 = 0$ that extends to small values of $\tan \beta$, and
two at large values of $|A_0|$ that are concentrated at larger $\tan
\beta$, particularly for $A_0 < 0$.  As seen in panel (b) of
Fig.~\ref{fig:gen}, these accumulations populate different regions of
$m_h$. Specifically, the points with $m_h \in (114, 119)$~GeV, which
populate the low-mass peak in Fig.~\ref{fig:mh}(b), have relatively low
values of $A_0$, most of which are negative. On the other hand, points
with $A_0 < - 2$~TeV generally have $m_h \in (119, 122)$~GeV and points
with $A_0 > 1$~TeV have $m_h \in (122, 127)$~GeV. Between these wings,
there are addditionally some low-$|A_0|$ points with $m_h \in (119,
124)$~GeV. Thus, the higher-mass peak in Fig.~\ref{fig:mh}(b) receives
contributions from all regions of $A_0$. We also see in
Fig.~\ref{fig:mh}(b) that essentially all the low-mass points are $(g_\mu
- 2)$-friendly (shaded red/grey), that only some of the high-mass points
with $A_0 > - 2$~TeV are $(g_\mu - 2)$-friendly, and that none of the
points with $A_0 < - 2$~TeV are $(g_\mu - 2)$-friendly.

\section{Interpretation of Features in the $m_h$ Distribution}

\begin{figure}
\vskip 0.5in
\vspace*{-0.75in}
\begin{minipage}{8in}
\epsfig{file=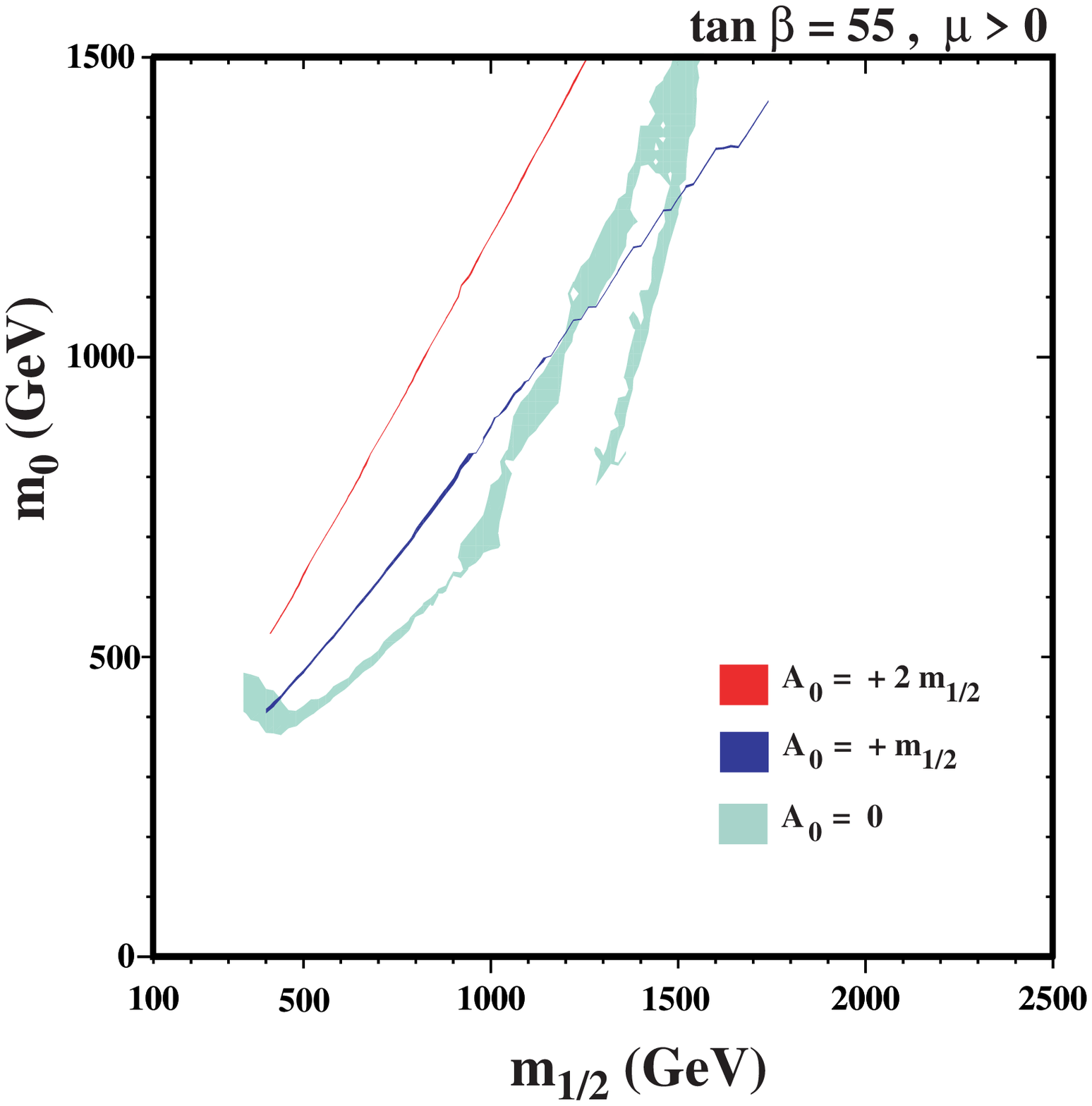,height=3.3in}
\hspace*{-0.17in}
\epsfig{file=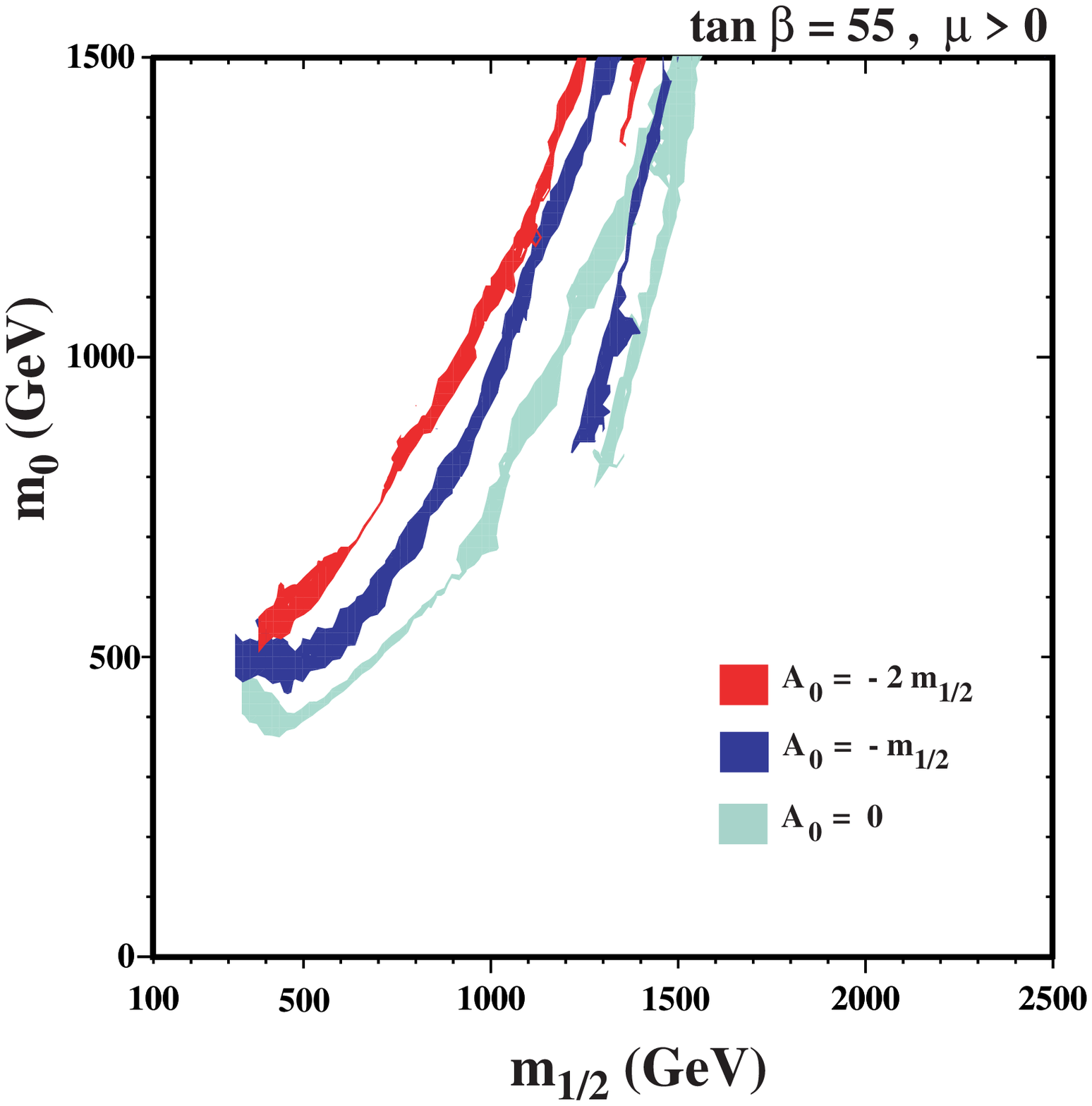,height=3.3in}
\hfill
\end{minipage}
\caption{
{\it
Regions in the $(m_{1/2}, m_0)$ planes for $\tan \beta = 55$, as 
calculated for $m_t = 174.3$~GeV using the latest version of the {\tt 
SSARD} code~\cite{SSARD}. Note the narrow coannihilation strips for $A_0 > 0$
in panel 
(a) and the broader rapid-annihilation funnels for $A_0 < 0$ shown in panel 
(b).}} 
\label{fig:ehow3} 
\end{figure}

The origins of many of these features can be understood qualitatively by
referring to the various $(m_{1/2}, m_0)$ planes displayed in Fig. 2
of~\cite{EHOW3} for different values of $\tan \beta$ and $A_0$. Updated
planes for the case $\tan \beta = 55$, whose importance can be seen from
panel (b) of Fig.~\ref{fig:tanbeta} and panel (a) of Fig.~\ref{fig:gen},
are shown in Fig.~\ref{fig:ehow3}. When $\tan \beta \la 45$ and $\mu > 0$
as assumed here, the regions allowed by WMAP and the other constraints are
essentially narrow {\it coannihilation strips} that decrease in width as
$m_{1/2}$ increases, terminating when $m_{1/2} \sim 1000$~GeV. In most of
these regions, $|A_0| \la 1000$~GeV also, so these points populate the
central $A_0 \sim 0$ region in Fig.~\ref{fig:gen}(b).  Therefore, they
provide the majority of the models in the {\it low-mass peak} in
Fig.~\ref{fig:mh}(b), but also a tail extending under the higher-mass
peak, as seen in panel (b) of Fig.~\ref{fig:gen}.  These coannihilation
strips are also the dominant features for $\tan \beta = 55$ when $A_0 >
0$, as seen in panel (a) of Fig.~\ref{fig:ehow3} by the two examples for
$A_0 = + m_{1/2}, + 2 m_{1/2}$.

However, a second class of features is visible in Fig. 2 of~\cite{EHOW3}
when $\tan \beta \ga 50$, namely {\it rapid-annihilation funnels} at large
$m_{1/2}$, as updated in panels (a, b) of
Fig.~\ref{fig:ehow3} for $A_0 = 0, - m_{1/2}, - 2 m_{1/2}$. These funnel
regions populate the {\it high-mass peak} in Fig.~\ref{fig:mh}(b). 
The funnels are typically broader than the coannihilation strips, and
therefore have a larger weighting in the constant-density sampling of the
$(m_{1/2}, m_0)$ plane that we have made in this paper. As a result, the
larger values of $\tan \beta$ have a strong weight in the sample of models
surviving the accelerator and WMAP constraints that we showed in
Fig.~\ref{fig:tanbeta}. We recall that we retain in our analysis points
whose relic density $\Omega_\chi h^2 $ falls below the range favoured by
WMAP~\cite{WMAP}, which typically have slightly lower values of $m_0$ than 
along the
coannihilation strips, while remaining within the region where the LSP is
the lightest neutralino $\chi$, and lie inside the rapid-annihilation
funnels. Restricting our plots to points with $\Omega_\chi$ within the
WMAP range would reduce the statistics in our plots, but not alter their
basic features. The weight of the rapid-annihilation points could be
diminished if one used a different sampling procedure, e.g., if one gave
less weight to regions of parameter space with large $m_{1/2}$ and/or
$m_0$, and hence $|A_0|$, as might be motivated by fine-tuning
considerations. However, the `twin-peak' structure of the $m_h$ 
distribution would survive any smooth reweighting of parameter space.

The rapid-annihilation funnels are responsible for the dense cluster of
models at large $\tan \beta$ and $A_0 < - 2$~TeV in panel (a) of
Fig.~\ref{fig:gen}, which have $m_h > 119$~GeV as seen in panel (b) of
Fig.~\ref{fig:gen}, and hence populate the higher peak in the Higgs mass
distribution in Fig.~\ref{fig:mh}. It is also clear from 
Figs.~\ref{fig:mh} and
\ref{fig:gen} that the basic feature of a doubly-peaked Higgs mass
distribution linked to different ranges of $A_0$ would also survive any
smooth reweighting of the parameter space.

As discussed in~\cite{EHOW3} and seen in Fig.~\ref{fig:ehow3}, the
locations of the rapid-annihilation funnels are very sensitive to $A_0$,
reflecting the sensitivity of $m_{A,H}$ to this parameter (among others).
Starting from the $A_0 = 0$ case where the funnel extends above $m_{1/2}
\sim 1000$~GeV for  $m_t = 174.3$~GeV, the funnel moves to smaller
$m_{1/2}$ as $A_0$ decreases and merges progressively with the
coannihilation strip. On the other hand, no funnels are visible for $A_0$
sufficiently $> 0$. The WMAP regions for $A_0 = \pm 2 m_{1/2}$ provide points in
Fig.~\ref{fig:gen} with extreme positive and negative values of $A_0$,
respectively. The different breadths of these regions explain the
asymmetry in panel (b) of Fig.~\ref{fig:gen}, in particular. Indeed, for
$A_0 > 0$ these points are simply continuations of the coannihilation
strips. As seen in Fig.~\ref{fig:gen}, some of these points are $(g_\mu -
2)$-friendly, and provide the tail under the high-mass peak in
Fig.~\ref{fig:mh}.

\begin{figure}
\vskip 0.5in
\vspace*{-0.75in}
\begin{minipage}{8in}
\epsfig{file=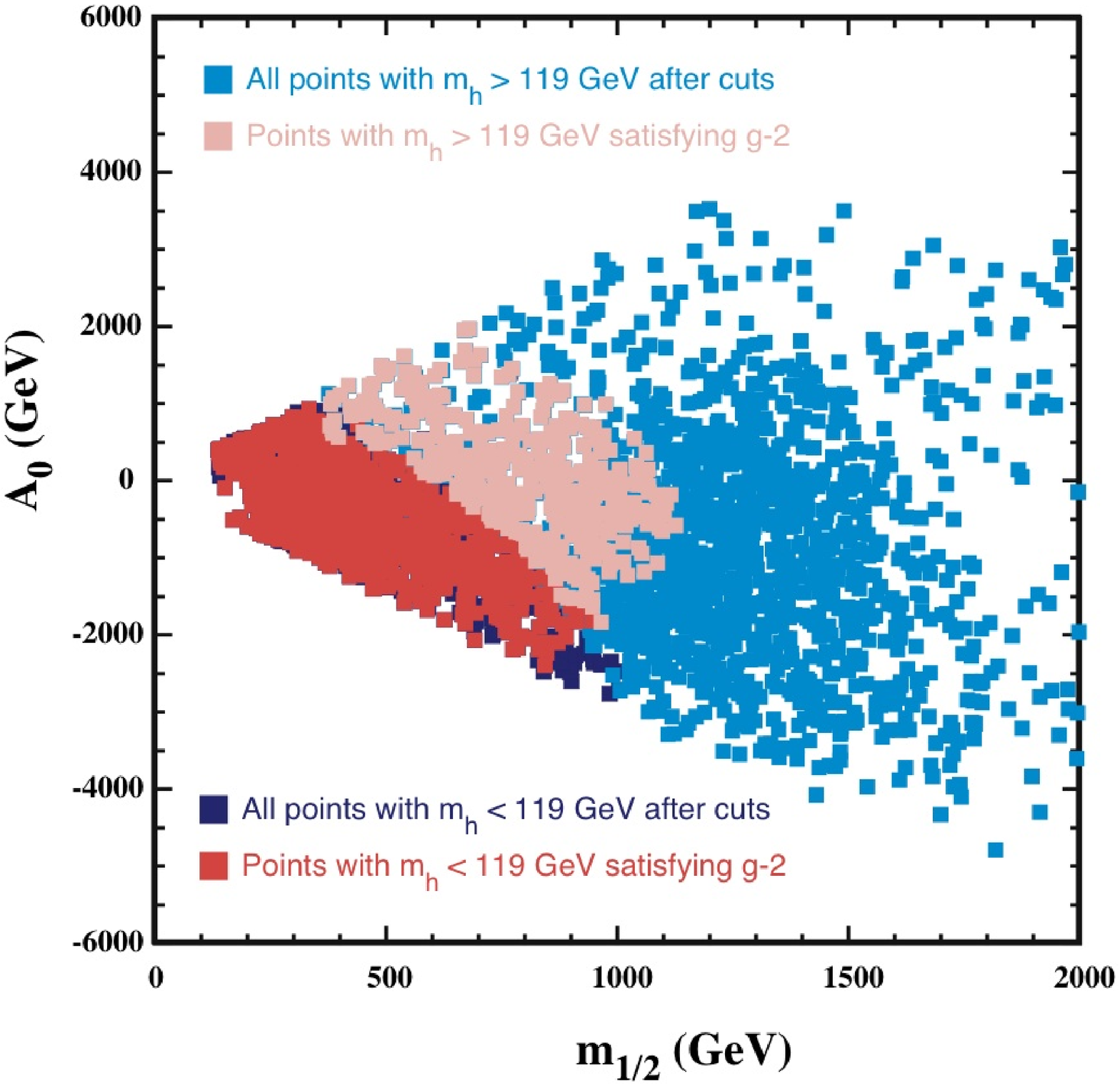,height=3.1in}
\hspace*{-0.17in}
\epsfig{file=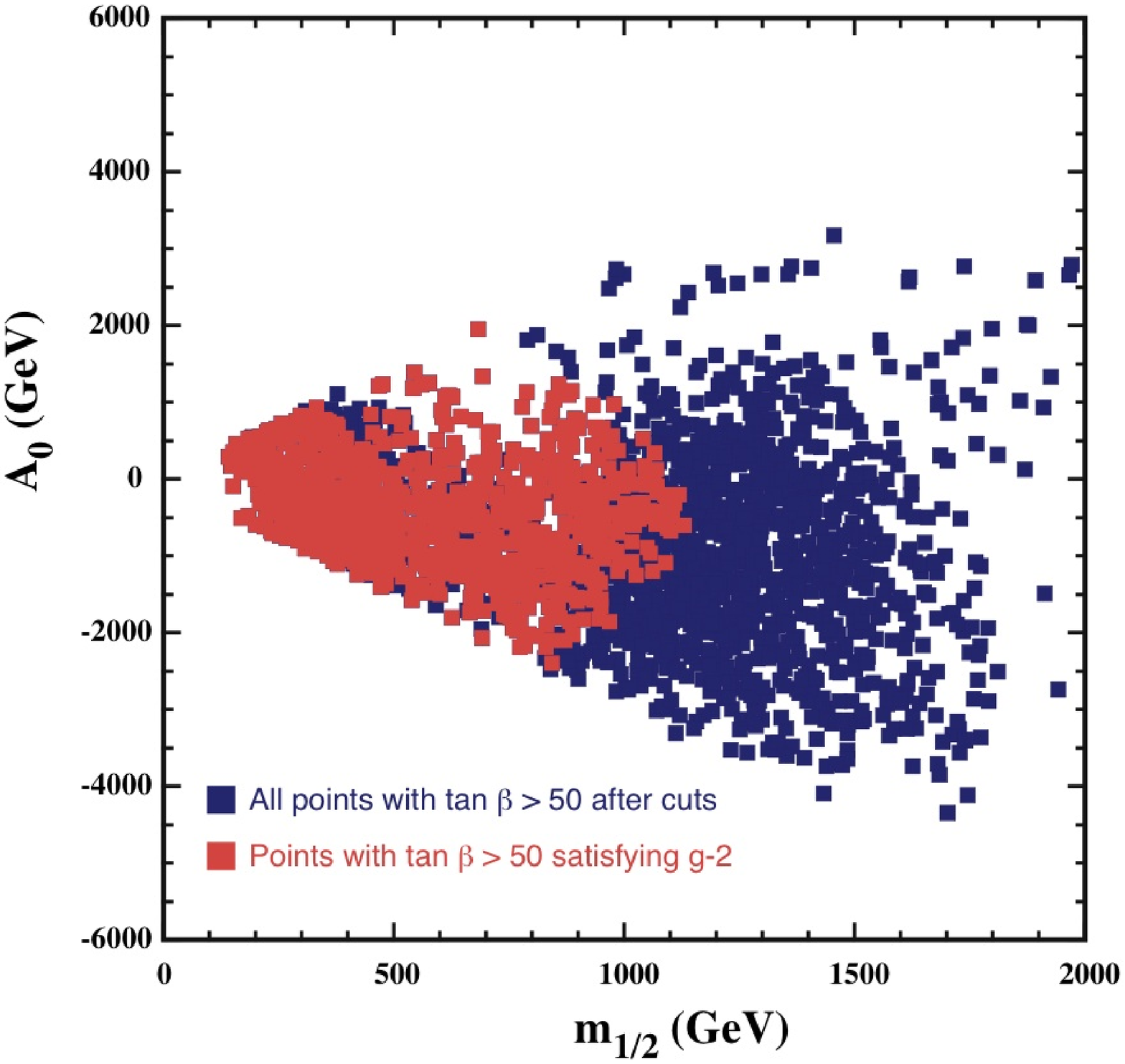,height=3.1in}
\hfill
\end{minipage}
\caption{
{\it
(a) The distribution in the $(m_{1/2}, A_0)$ plane of the 
points shown in 
Fig.~\protect\ref{fig:mh}(b) that satisfy the accelerator and WMAP 
constraints, separated into the low-$m_h$ region (blue points) and 
the high-$m_h$ region (red points). (b) The distribution in the $(m_{1/2}, 
A_0)$ plane of points with $\tan \beta > 50$.}} 
\label{fig:fish}
\end{figure}

We now consider the $(m_{1/2}, A_0)$ planes shown in Fig.~\ref{fig:fish},
where panel (a) shows the combination of all vales of $\tan \beta$, and
panel (b) shows only models with $\tan \beta > 50$.  These panels update
analogous plots in~\cite{EHOW3}, and display significant differences due
to the reduction in $m_t$ from 178~GeV to 174.3~GeV and improvements in
the treatment of vacuum stability requirements. Previously, we had seen a
clear separation between `fins' at $A_0 \sim \pm 1.5 m_{1/2}$ and a
`torso' at $A_0 \sim 0$, which has vanished apart (possibly) from a
vestigial ´fin´ at $A_0 > 2$~TeV that is more visible at large $\tan
\beta$.  We also note the apüpearance of a small `head' with a `tooth' at
$m_{1/2} \la 150$~GeV and $A_0 \sim 0$, which is due to points with
$m_\chi \sim m_h/2$, whose relic density falls within the WMAP range
thanks to rapid annihilation through the light CMSSM Higgs 
pole~\cite{Drees}. These
points have $m_{\chi^\pm}$ very close to the LEP lower limit, and might be
accessible to the Tevatron.

We see in panel (a) of Fig.~\ref{fig:fish}
that the points with $m_h < 119$~GeV (darker blue/black and 
red/grey colours) cluster at $m_{1/2},
A_0 < 1$~TeV and $A_0 > - 3$~TeV. Almost all these points make a
supersymmetric contribution to $g_\mu - 2$ that could explain the possible
discrepancy between experiment and the Standard Model calculation based on
$e^+ e^-$ data (indicated in red/grey). On the other hand, only a small
fraction of the models with $m_h > 119$~GeV (pale colours) are compatible
with this supersymmetric interpretation of $g_\mu - 2$ (pink/light grey). 
As seen in
panel (b) of Fig.~\ref{fig:fish}, all the parameter sets with $\tan \beta
> 50$ have $m_h > 119$~GeV. The $(g_\mu - 2)$-friendly points are 
concentrated at $m_{1/2} \lappeq 1$~TeV.

This analysis can be used as a diagnostic tool when the Higgs boson is
discovered at the Tevatron or the LHC, at least within the CMSSM framework 
and assuming
that $m_t = 174.3$~GeV. This framework would be invalidated if $m_h >
127$~GeV. On the other hand, if the Higgs boson is discovered with a mass
$m_h < 119$~GeV, one can infer from Fig.~\ref{fig:fish}(a) that both
$m_{1/2}$ and $A_0$ must be small, and that supersymmetry is likely to 
lie along a coannihilation strip. On the other hand, if $m_h > 119$~GeV, 
supersymmetry may well have chosen a rapid-annihilation funnel.

\section{Potential Impact of $g_\mu - 2$}

We now comment further on the potential impact of imposing the $g_\mu - 2$
constraint~\cite{g-2}, which we treat as optional. We see in 
Fig.~\ref{fig:mh}(b)  
that this constraint would suppress the high-$m_h$ peak, while retaining
most of the low-$m_h$ models. The suppression of the high-$m_h$ peak is a
consequence of the removal of points with large $m_{1/2}$ and/or $m_0$
that would make a very small contribution to $g_\mu - 2$, many of which
are in the rapid-annihilation funnels. A similar effect reduces also the
upper part of the low-$m_h$ peak, but the coannihilation strips would be
less affected by the $g_\mu - 2$ constraint. On the other hand, as seen in
Fig.~\ref{fig:tanbeta}, imposing the $g_\mu - 2$ constraint would not
alter the statistical preference for large $\tan \beta$. As we see in
Fig.~\ref{fig:gen}, imposing the $g_\mu - 2$ constraint would disfavour
models with large negative $A_0$, as well as many with large positive
$A_0$, but some models with large $\tan \beta$ and a small $A_0$ would
survive.

\section{Dependence on $m_t$}

\begin{figure}
\vskip 0.5in
\vspace*{-0.75in}
\begin{minipage}{8in}
\epsfig{file=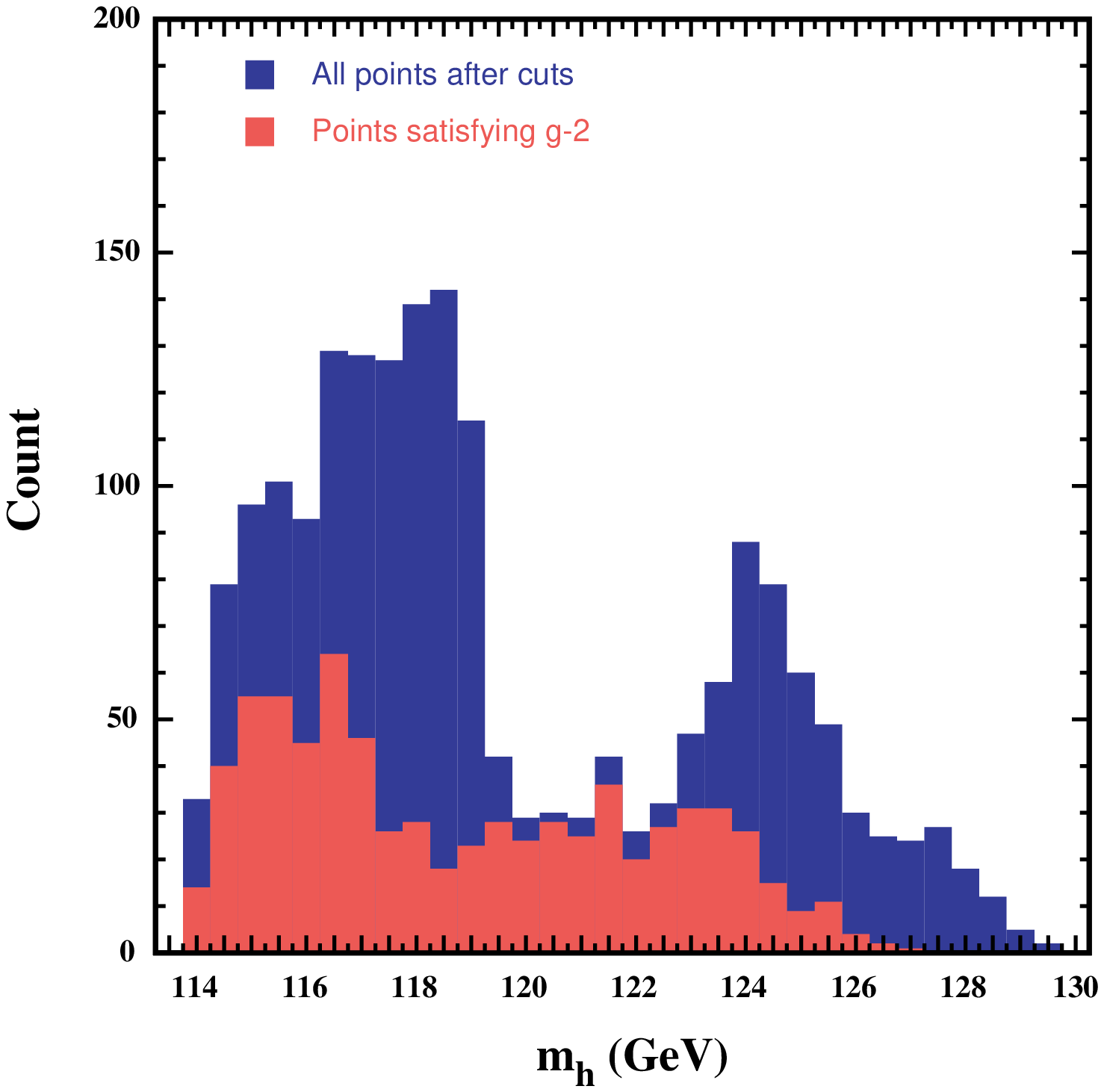,height=3.1in}
\epsfig{file=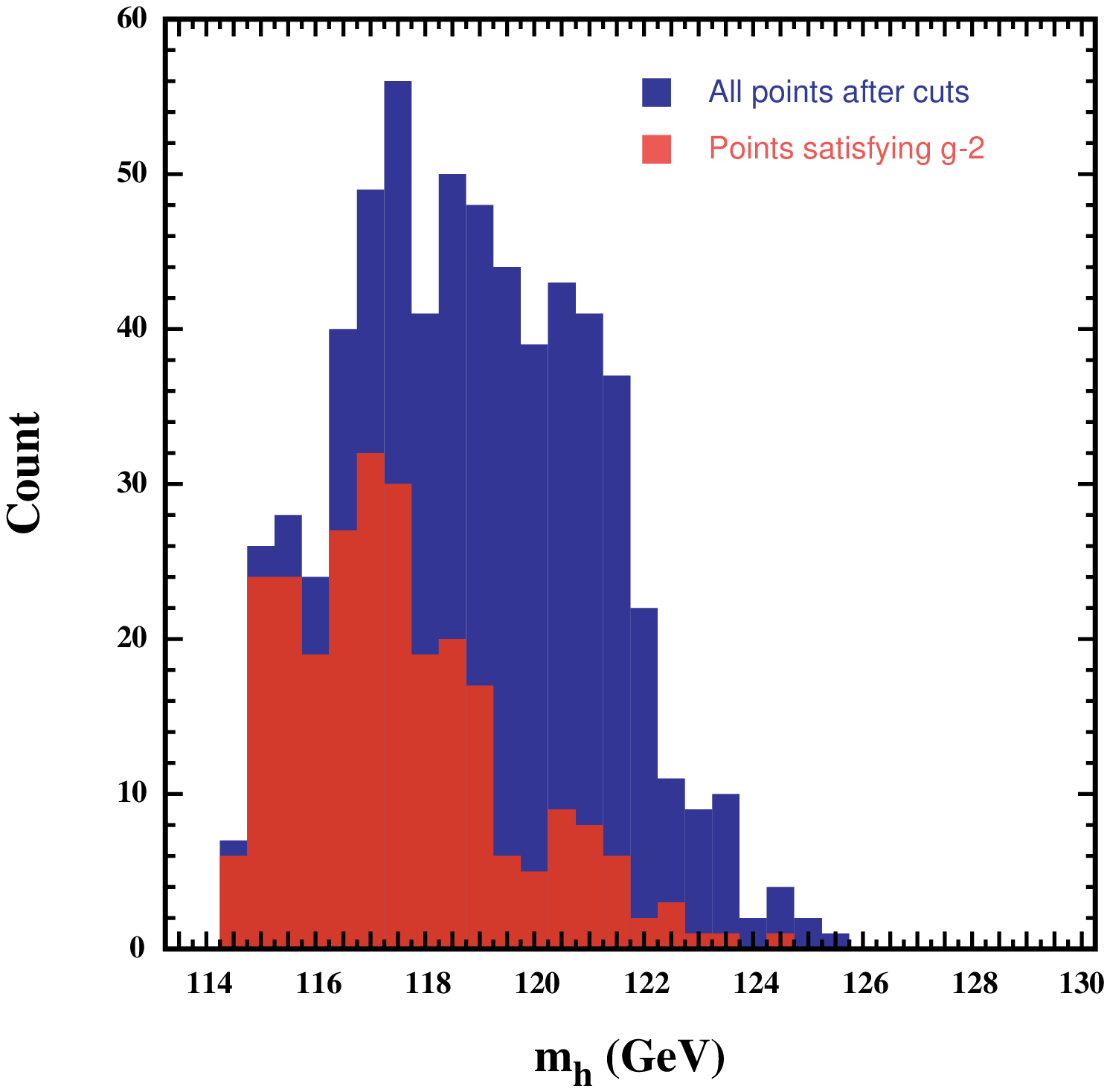,height=3.1in}
\end{minipage}
\caption{
{\it
As for panel (b) of Fig.~\protect\ref{fig:mh}, but assuming (a) $m_t = 
178$~GeV and (b) $m_t = 172.7$~GeV.
}} 
\label{fig:mt}
\end{figure}

In all the above, we have assumed that $m_t =174.3$~GeV~\cite{mtop174}. 
The central value was formerly 178~GeV~\cite{mtop178}, and the
current central value is $m_t = 172.7 \pm 2.9$~GeV~\cite{mtop172}, 
following significant evolution during recent months. In view of this and 
the remaining
experimental uncertainty, we have also considered the dependence of the above
analysis on $m_t$. We recall that $\Delta m_h \sim 1$~GeV for $\Delta m_t =
1$~GeV in theoretical calculations, and that the parameter regions allowed by
WMAP vary quite considerably with $m_t$, particularly in the 
rapid-annihilation
funnel region, as seen in Fig.~1 of~\cite{EHOW3}. Specifically, this region
moves to smaller (larger)   $m_{1/2}$ for smaller (larger) $m_t$. As was already
mentioned, the coannihilation strips mainly populate the low-mass peak in
Fig.~\ref{fig:mh} whereas the high-mass peak is largely due to the
rapid-annihilation funnels. Accordingly, we would expect these peaks to be more
separated at large $m_t$ than at smaller values. Precisely this effect is seen
in the two panels of Fig.~\ref{fig:mt}. We see in panel (a)   that the upper
peak in Fig.~\ref{fig:mh}(b) shifts upwards by $\sim 4$~GeV if $m_t =
178$~GeV~\cite{mtop178}, and is very clearly separated from the low-mass peak.  
Correspondingly, the upper limit on $m_h$ increases to 130~GeV. On the other
hand, we see in panel (b) of Fig.~\ref{fig:mt} that the two peaks merge for $m_t
= 172.7$~GeV~\cite{mtop172}, and the upper limit on $m_h$ decreases to
126~GeV.   Likewise, many of the other features discussed previously in this
paper become more (less) pronounced for larger (smaller) $m_t$.

By the time the Higgs boson is discovered, we expect $m_t$ to be known
with considerably greater precision than the present uncertainty $\delta
m_t \simeq 2.9$~GeV. Once $m_t$ is known with an accuracy $\delta m_t \la
2$~GeV, and assuming that the accuracy of theoretical calculations in the
CMSSM can be brought to the same level, there will be no theoretical
ambiguity between the Higgs mass peaks, and diagnosis of the
supersymmetric parameters will indeed be possible along the lines
discussed in the previous Section.

As in the case $m_t = 174.3$~GeV shown in panel (a) of Fig.~\ref{fig:mh},
for $m_t = 178$~GeV the effect of imposing the $g_\mu - 2$ constraint
would also be to remove the high-mass peak, leaving a plateau extending
from $m_t \sim 118$~GeV to $\sim 127$~GeV. The low-mass-peak would also be
reduced, but most of the intermediate plateau for $m_t = 178$~GeV would
survive the $g_\mu - 2$ constraint. In the case of $m_t = 172.7$~GeV shown
in panel (b), there is a more pronounced peak at $m_t \sim 117$~GeV and a
tail extending to $\sim 124$~GeV. It is striking that, whatever the value
of $m_t$, the mode of the $m_h$ distribution is relatively stable at $\sim
117$~GeV and that the upper limit on $m_h$ also remains relatively stable
around 125~GeV, if the $g_\mu - 2$ constraint is imposed.

\section{Conclusions}

We have shown that the available experimental and cosmological constraints
on the CMSSM allow only a limited range of $m_h$. If $m_t = 174.3$~GeV,
this is $< 127$~GeV without the $g_\mu - 2$ constraint and $< 124$~GeV if
it is imposed. If $g_\mu - 2$ is not imposed, we find twin peaks in the
$m_h$ distribution at $m_h \sim 117, 121$~GeV. The upper bound and the
lower peak are quite insensitive to variations in $m_t$, whereas the upper
peak is sensitive, and merges with the lower peak for low $m_t$. Large
values of $\tan \beta \sim 55$ are favoured by our analysis, whether the
$g_\mu - 2$ constraint is applied, or not.

We have also shown in this paper that the mass of the lightest CMSSM Higgs
boson may be a useful diagnostic tool for identifying the most likely
regions of the CMSSM parameter space, even if sparticles are not (yet)  
discovered. This is because the CMSSM is divided into distinct
coannihilation and rapid-annihilation regions, and measuring $m_h$ could
provide us with a hint which alternative is chosen by Nature.

\section*{Acknowledgments}
\noindent 
The work of D.V.N. was supported in part
by DOE grant :DE-FG03-95-ER-40917.
The work of K.A.O. was supported in part
by DOE grant DE--FG02--94ER--40823.
The work of Y.S. was supported in part by
the NSERC of Canada, and Y.S. thanks the Perimeter Institute for its
hospitality.

\end{document}